# Realizing tight-binding Hamiltonians using site-controlled coupled cavity arrays

**Authors:** Abhi Saxena[1]*, Arnab Manna[2], Rahul Trivedi[1] and Arka Majumdar[1,2]*

**Affiliations:**
[1]Department of Electrical & Computer Engineering, University of Washington; Seattle, Washington, 98195, USA.
[2]Department of Physics, University of Washington; Seattle, Washington, 98195, USA.

*Correspondence to: <u>abhi15@uw.edu</u>, <u>arka@uw.edu</u>

**Abstract:** Analog quantum simulators rely on programmable quantum devices to emulate Hamiltonians describing various physical phenomenon. Photonic coupled cavity arrays are a promising platform for realizing such devices. Using a silicon photonic coupled cavity array made up of 8 high quality-factor resonators and equipped with specially designed thermo-optic island heaters for independent control of cavities, we demonstrate a programmable device implementing tight-binding Hamiltonians with access to the full eigen-energy spectrum. We report a ~50% reduction in the thermal crosstalk between neighboring sites of the cavity array compared to traditional heaters, and then present a control scheme to program the cavity array to a given tight-binding Hamiltonian.

**One-Sentence Summary:** We report a programmable photonic cavity array to implement tight binding Hamiltonian with access to the full eigen-energy spectrum.

Achieving analog quantum simulation necessitates the realization of programmable quantum devices (*1*). Due to their inherent driven-dissipative nature, photonic systems are a promising platform for non-equilibrium quantum simulation (*2*). An archetypal photonic quantum simulator consists of an array of programmable non-linear nodes with access to the entire quantized eigen-energy spectra of the Hamiltonians being simulated. While there have been numerous works on analog quantum simulation with microwave photons (*3–6*), higher energy optical photons can provide several additional advantages, including operability at much higher temperatures (*7*), which would significantly simplify the experiments and lower the resources needed to scale the simulator; and availability of single photon detectors, which would allow direct measurement of multiparticle correlations (*8, 9*).

One solution to engineer such systems in optics is via photonic coupled cavity arrays (CCA) (*10*) where coupling between cavities provides a potential map for photons to move around, and strong spatial confinement of light for long durations allows access to onsite non-linearity via coupling with various excitonic materials. The latter demand, in practice, translates to using high-quality factor (Q) cavities with small mode volumes as constituents of the CCA. Though several experiments showing various physical phenomena using optical CCAs have been previously reported (*11–13*), none of these CCAs are programmable and can have access to the entire quantized eigen-energy spectra of the Hamiltonian. In the optical regime achieving these capabilities, namely programmability and measurability of the eigen-spectrum, is very challenging owing to the extremely small physical dimensions involved.

In this work, we tackle these problems by engineering a silicon photonic CCA made of high Q ($\sim 8.5 \times 10^4$) racetrack resonators with thermally controllable onsite potential using specially designed thermo-optic (TO) island heaters. Here, we specifically focus on 1D tight-binding lattices which can be described by the set of Gaussian Hamiltonians of the form ($\hbar=1$):

$$H = \sum_n \mu_n a_n^\dagger a_n + J_n(a_{n+1}^\dagger a_n + a_n^\dagger a_{n+1}) \tag{1}$$

where $a_n$ denote the onsite photonic annihilation operator, $\mu_n$ is the onsite potential given by the resonant frequency of the cavity, $J_n$ is the photonic hopping rate between $n^{th}$ and $(n+1)^{th}$ sites. Realization of such a set of Hamiltonians requires implementing a potential profile $[\mu_n] = [\mu_0, \mu_1, \ldots \mu_{N-1}]$ across a photonic lattice with specific inter-site hopping rates $J_n$, while ensuring that all the eigenstates of the system denoted by $[\epsilon_n] = [\epsilon_0, \epsilon_1, \ldots \epsilon_{N-1}]$ remain addressable and measurable.

Experimentally, we implement a Hamiltonian with 8 nodes via a CCA made up of 8 strongly coupled racetrack resonators fabricated on a silicon-on-insulator platform using $220\ nm$ silicon on top of $3\ \mu m$ thick silicon oxide (Fig. 1A). The spacing between the resonators is determined by the desired hopping rate between the sites for the tight binding Hamiltonians being implemented. The spectrum of the resulting system is probed via a set of grating couplers located at the first and last sites. The scattering properties of this system are completely described by the effective non-Hermitian Hamiltonian which incorporates the coupling to input/output ports and system losses as:

$$H_{eff}^0 = H - j\left(\frac{\gamma_0}{2} a_0^\dagger a_0 + \frac{\gamma_{N-1}}{2} a_{N-1}^\dagger a_{N-1}\right) - j \sum_n \frac{\kappa_n}{2} a_n^\dagger a_n \tag{2}$$

where $\gamma_0, \gamma_{N-1}$ denote the coupling rates to the grating couplers and $\kappa_n$ denotes the onsite scattering/absorption losses. To map this initial Hamiltonian $H_{eff}^0$, we extend the Hamiltonian tomography algorithm developed for $1D$ lossless lattices (*14, 15*) for application in $1D$ nearest neighbor lossy CCAs (see supplementary information). The modified algorithm allows for determining the entire $H_{eff}^0$ describing the system from a single reflection spectrum measurement $|R(\omega)|^2$ (Fig. 1B) by estimating the contribution of individual eigenmodes of the system to the measured spectrum. In Fig. 1C, we plot the reflection spectrum of our CCA along with corresponding contributions of the 8 individual eigenmodes. We then verify the accuracy of our fit by comparing the experimentally measured transmission spectrum $|T(\omega)|^2$ of the CCA to the predicted spectrum of the extracted $H_{eff}^0$. Note that, while the reflection spectrum is needed to map the entire $H_{eff}^0$, the transmission spectrum can be used to find only the eigenvalues of the Hamiltonian (see supplementary information).

Thermal control of the CCA has are two primary design objectives: (i) minimizing the additional optical loss incurred when introducing the heaters, and (ii) reducing the thermal crosstalk between heaters which need to be placed in close proximity owing to the small device footprint necessary to obtain small mode volumes for each cavity and ensure strong coupling between the cavities (*16*). We meet both objectives by engineering TO island heaters made up of tungsten ($W$) wires sandwiched between two alumina ($Al_2O_3$) layers (Fig. 2A, B). In such a configuration the lower thermal resistance of the alumina layers than that of the air/silicon oxide channel separating the islands allows for a more directional transfer of thermal energy from the tungsten heaters to the corresponding resonators. Our approach succeeds in curbing the effects of thermal crosstalk between neighboring resonator sites by ~ 50% compared to typically used TO heaters (*17, 18*) (see supplementary information). Since, alumina is optically lossless in the telecommunication wavelength range, the islands also allow for placing the tungsten heaters at an adequate distance from the racetrack resonators. This ensures that the introduction of heating elements occurs with minimal absorptive losses. Overall, this mode of operation allows achieving much higher Q-factors required for addressability of individual eigenmodes of a controllable CCA platform than previously reported approach involving photoconductive elements (*19*).

We characterize the CCA by applying a linearly increasing voltage across each heater one at a time and recording the respective transmission spectra. The eigen-energies are then extracted from the recorded spectra and combined with our knowledge of $H_{eff}^0$, we estimate the amount of crosstalk between the heaters. The change in the onsite potential $\Delta\mu_n$ when expressed in wavelength units is proportional to the square of voltage $V_n$ applied to the $n^{th}$ site: $\Delta\mu_n^\lambda \propto V_n^2$ (see supplementary information). To simplify the equations going forward, we express the onsite potentials $\mu_n$ and eigenvalues $\epsilon_n$ of the CCA in wavelength units as $\mu_n^\lambda$ and $\epsilon_n^\lambda$. We plot the effects of voltage $V_n$ applied across heater $h_n$ on the potential profile $[\mu_n^\lambda]$ of the CCA in Fig. 2C. The change in respective onsite potentials $\Delta\mu_n^\lambda$ is represented by the radii of the circles, whereas the color of the circles denotes the voltage $V_n$ applied across heater $h_n$. From the plot, we establish that thermal crosstalk is already low between nearest neighbors ($n \pm 1$) and becomes negligible as we go beyond third nearest neighbors ($n \pm 3$).

We next model the CCA to accurately predict the eigen-energies of the system on application of a voltage profile $[V_n] = [V_0, V_1, ... V_{N-1}]$ across the heaters. Based on the observation that thermal crosstalk is restricted to three nearest neighbors, we expect the change in onsite potential $\Delta\mu_n^\lambda$ at the $n^{th}$ site on application of $[V_n]$ to depend on $V_i^2$ and their cross products $V_j V_k$ s.t. $i, j, k \in [n-3, n+3]$. We can express this relation mathematically through a function $f$ as

$$\Delta \mu_n^\lambda = f([V_n]) = \delta_n + \sum_i \beta_i (\alpha_i V_i^2) + \sum_{j,k} \gamma_{j,k} (\sqrt{\alpha_j \alpha_k} V_j V_k) \tag{3}$$

where $\delta_n$ is a fitting correction to the initial onsite potential, $\beta_i$'s are proportionality coefficients for the direct terms, $\gamma_{j,k}$'s are the proportionality coefficients for the cross-terms and coefficients $\alpha_n$ are used to incorporate the effects of minor variations in heater resistances due to fabrication inconsistencies. Note that, owing to the geometrical symmetry of the CCA about any $n^{th}$ site, we assume that the number of functional parameters $\beta_i$ and $\gamma_{j,k}$ needed to model the device behavior can be restricted to 3 and 12 respectively.

We visualize this process in Fig. 3A where we show how we can use the model to predict the location of eigen-energies $[\epsilon_n^\lambda]$ by finding the eigenvalues of the modified Hamiltonian. Starting with the initial $H_{eff}^0$ and updating its diagonal terms by evaluating the function $f$ at each site of the array for a particular $[V_n]$ we predict the eigenvalues of the modified Hamiltonian as:

$$[\epsilon_n^\lambda]_{predicted} = Eig(H_{eff}^0 + [\Delta \mu_n] \mathbb{I}) \tag{4}$$

These predicted eigen-energies are then used to fit for $f$ by minimizing the error obtained by calculating the deviations from experimentally extracted eigen-energies across many measurements (here we limit the number of measurements to 288).

$$Error = \frac{\| [\epsilon_n^\lambda]_{predicted} - [\epsilon_n^\lambda]_{measured} \|^2}{J_{norm}} \tag{5}$$

The probability distribution of the fitting error normalized to the mean hopping-rate $J_{norm}$ is plotted in Fig. 3B. Finally, once we have identified $f$, we use it to predict the location of eigen-energies for 20 randomly generated voltage profiles in Fig. 3C. The centers of the circles in the figure denote the measured values of eigen-energies, and the error in predicted values are represented by the radii of the corresponding circle. The net overall error for a random generation is mapped to the color of the particular set of eigen-energies. From the plot we can see that the model allows for prediction of the eigen-energies of our system with greater than 96% accuracy.

We demonstrated a thermally controlled optical CCA which can be used to realize a set of tight binding Hamiltonians with addressability to the entire quantized eigen-energy spectrum. To ensure a compact device footprint and high Q cavities necessary for reaching the regime of interacting photons (*10*), we engineered special TO islands heaters, which allowed reduction in thermal crosstalk by almost 50% over previously reported works (*17, 18*). Finally, we present a mathematical model which allowed for precise control of the eigen-energies of the implemented Hamiltonians within an error of only 4% of the mean hopping rate. While we did not demonstrate any non-linearity, our CCA with high Q-factors and a cladding free design will readily allow integration with excitonic materials and possibly reaching single photon non-linearities (*20*).

**Acknowledgments:**

Part of this work was conducted at the Washington Nanofabrication Facility / Molecular Analysis Facility, a National Nanotechnology Coordinated Infrastructure (NNCI) site at the University of Washington with partial support from the National Science Foundation via awards NNCI-1542101 and NNCI-2025489. We thank J. Simon for pointing us towards the Hamiltonian tomography algorithms; M. Zhelyeznyakov and S.L. Brunton for help in numerical optimization; J. Whitehead for help in automating the measurements; Y. Chen for help in fabrication in initial stages of the project.

**Funding:**

National Science Foundation grant NSF-QII-TAQS-1936100

**Competing interests:** Authors declare that they have no competing interests.

**Data and materials availability:** All data needed to evaluate the conclusions in the paper are present in the paper and/or the Supplementary Materials. Additional data related to this paper may be requested from the authors on reasonable request.

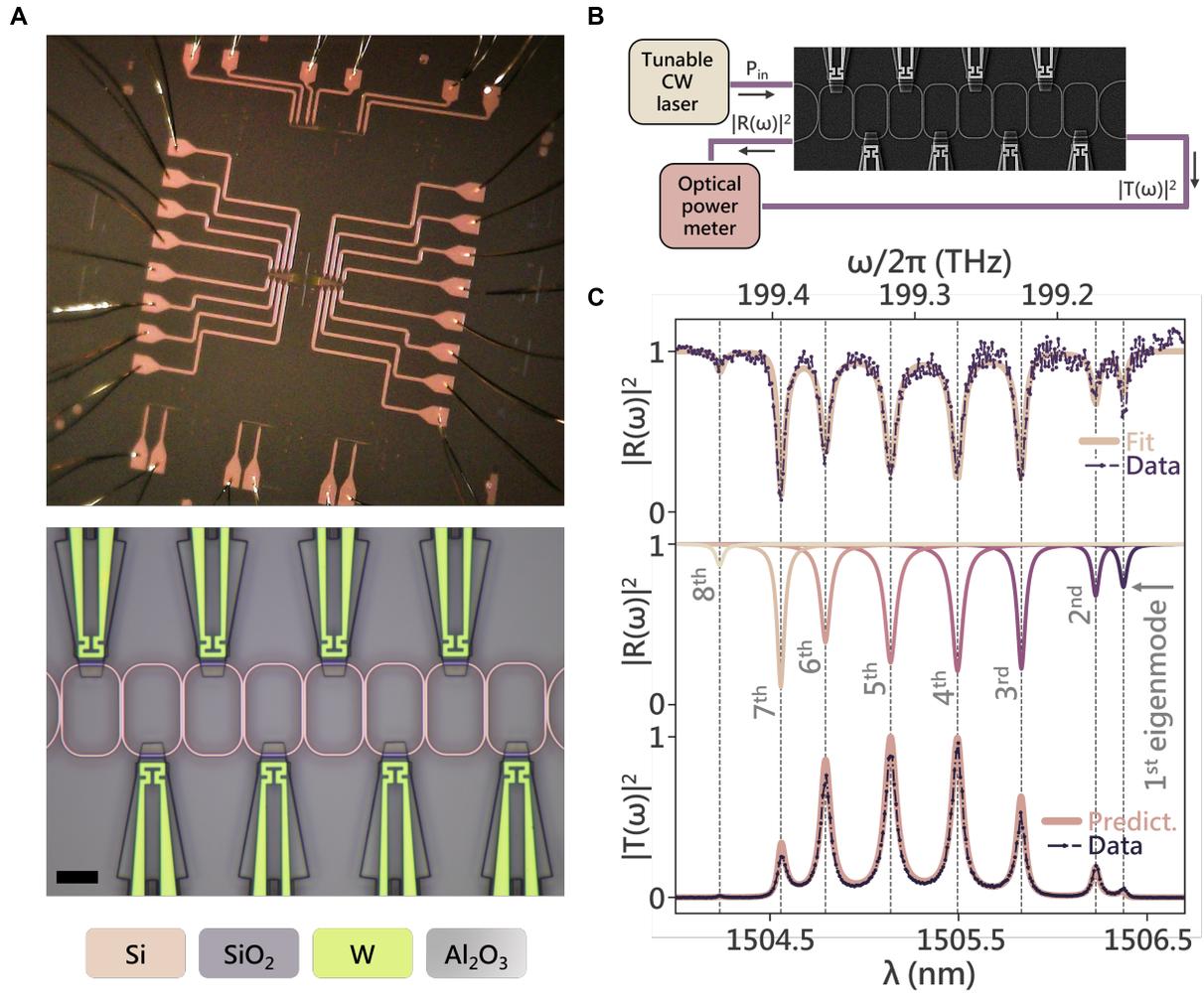

**Fig. 1 Hamiltonian Tomography.** (A) Optical image of the electrically controlled CCA depicting the wiring structure, optical micrograph of the CCA (scale bar: $10\ \mu m$). (B) Schematic of the experimental setup used for measuring reflection ($|R(\omega)|^2$) and transmission ($|T(\omega)|^2$) spectra. (C) From the top: measured reflection spectrum $|R(\omega)|^2$ (dotted purple) along with the fit generated using the tomography algorithm (cream); followed by a plot showing contributions of various eigenmodes of the system to $|R(\omega)|^2$, and finally at the bottom; experimentally measured transmission spectrum $|T(\omega)|^2$ (dotted purple) along with the predicted transmission spectrum $|T(\omega)|^2$ (pink) from the $H_{eff}^0$ obtained using tomography algorithm.

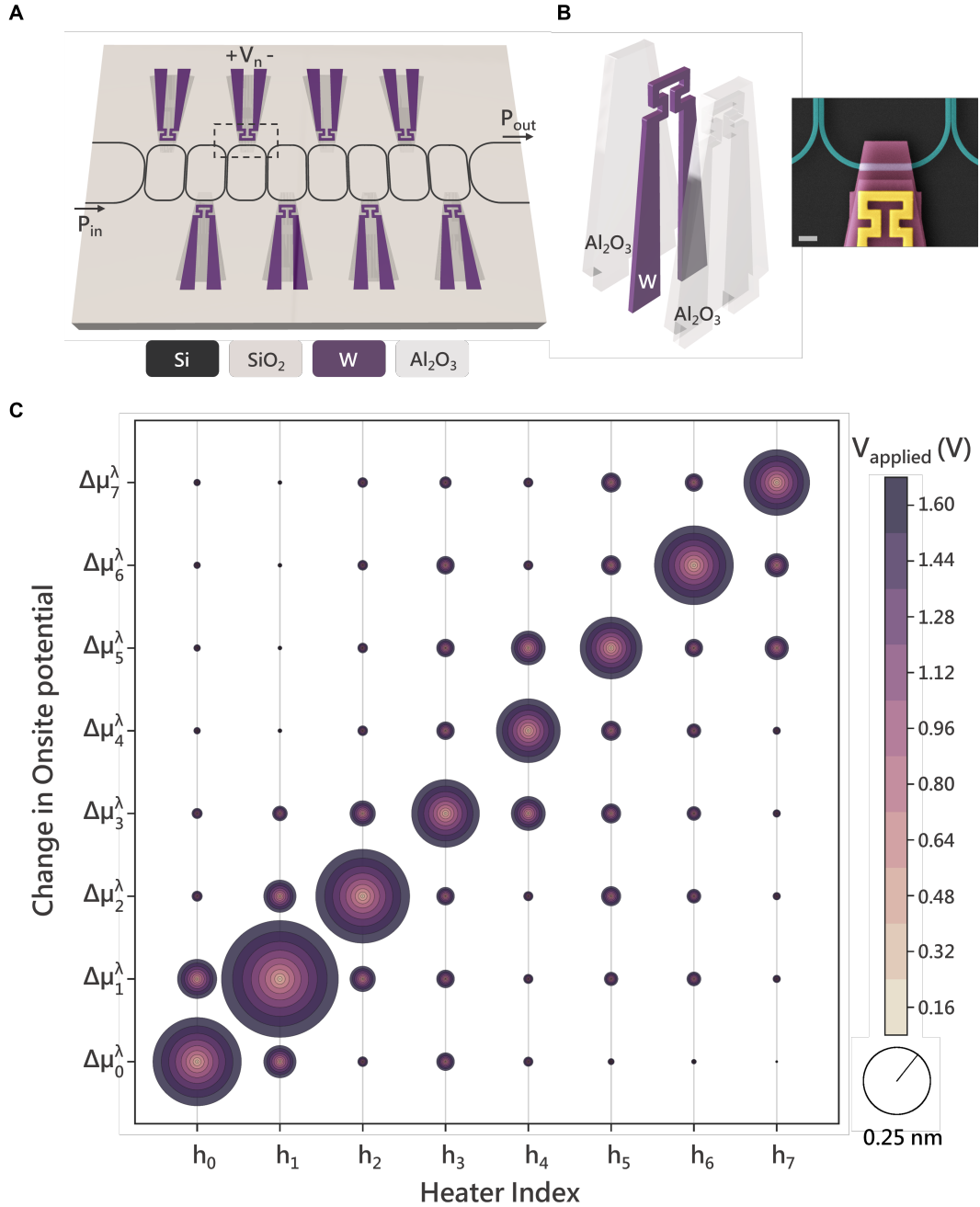

**Fig. 2 Electrical Characterization.** (A) Device schematic depicting the electrical characterization as voltage $V_n$ is applied to the $n^{th}$ site while measuring the transmission spectrum. (B) Exploded view of the TO island heaters. The heater consists of a tungsten element sandwiched between alumina layers. Inset shows a false colored SEM image (scale bar: $2\ \mu m$) of a typical TO island (yellow: tungsten, pink: alumina, teal: silicon). (C) Plot showing the effect of heaters [$h_n$] on the potential profile across the device. The x-axis denotes the heater index $h_n$ switched ON for a particular set of measurements and the y-axis represents the change in potential profile [$\Delta\mu_n^\lambda$]. The voltage applied of the measurement $V_n$ across heater $h_n$ is mapped to the color of the circular surface and the corresponding change in potential is denoted by the radii of the circle encompassing the surface ($0.25\ nm$ of change is depicted by radii of the circle in the scale bar).

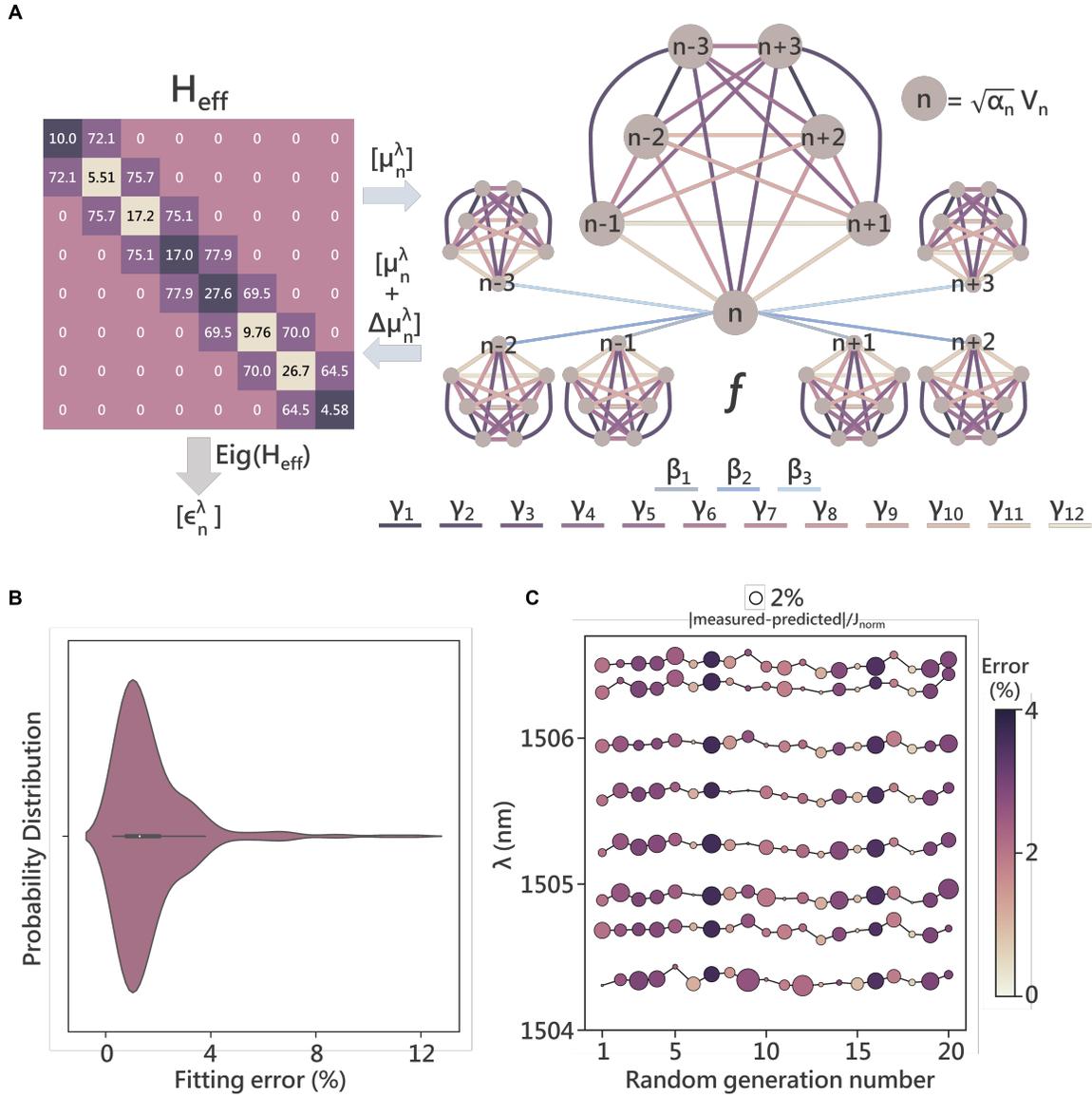

**Fig. 3 Electrical control model and eigen-energy prediction.** (A) Visualization of the optimization process depicting how the model takes in the system Hamiltonian $H_{eff}$ and fits for the function $f$ which connects the applied voltage profile $[V_n]$ to change in onsite potentials. We predict the position of eigen-energies on application of $[V_n]$ by calculating the change in onsite potentials which lie along the Hamiltonian diagonal and finding the eigenvalues of the modified Hamiltonian. The optimization is initialized using $H_{eff}^0$ shown in the matrix form (only real part is depicted). All entries are in $GHz$, with diagonal terms denoting the deviations in resonant frequency about the mean (dark purple: $+ve$ deviation, tan: $-ve$ deviation) and super/sub diagonal terms denoting the hopping rates. The coefficients $\beta_i$ are denoted in hues of blue and $\gamma_{j,k}$ are denoted in hues of purple. (B) Violin plot denoting the fitting error normalized to the mean hopping rate $J_{norm}$ across 288 points. (C) Prediction accuracy plot where the x-axis denotes the random generation and the y-axis denotes the wavelength. The location of the measured eigen-energies is denoted by the dark black lines in background. The radii of the circles denote the deviation of the predicted value from measured values (scale bar on top). The color of the dots denotes the overall prediction error for that generation.

## Materials and Methods

**Design:**

Ansys Lumerical FDTD and HEAT were used to simulate and optimize the device parameters.

**Fabrication:**

A silicon on insulator wafer (SOITEC) with $220\ nm$ thick film of silicon on $3\ \mu m$ thick buried silicon oxide ($BOX$) was cleaved. A $10\ mm \times 10\ mm$ chip thus obtained was used for further processing. The chip was cleaned by sonication in acetone and isopropyl alcohol followed by an oxygen plasma. The chip was then spin-coated with $HSQ$ and exposed using a JEOL $JBX6300FS$ electron beam lithography system. After developing in 25% $TMAH$, the chip was etched using an inductively coupled plasma etcher with a $Cl_2$ chemistry. The resist was then removed using diluted $BOE$. The chip then underwent several cycles of electron beam lithography based $PMMA$ patterning, followed by electron beam evaporation/sputtering of materials and liftoff to define the island heaters and contact pads as depicted in Fig. S1.

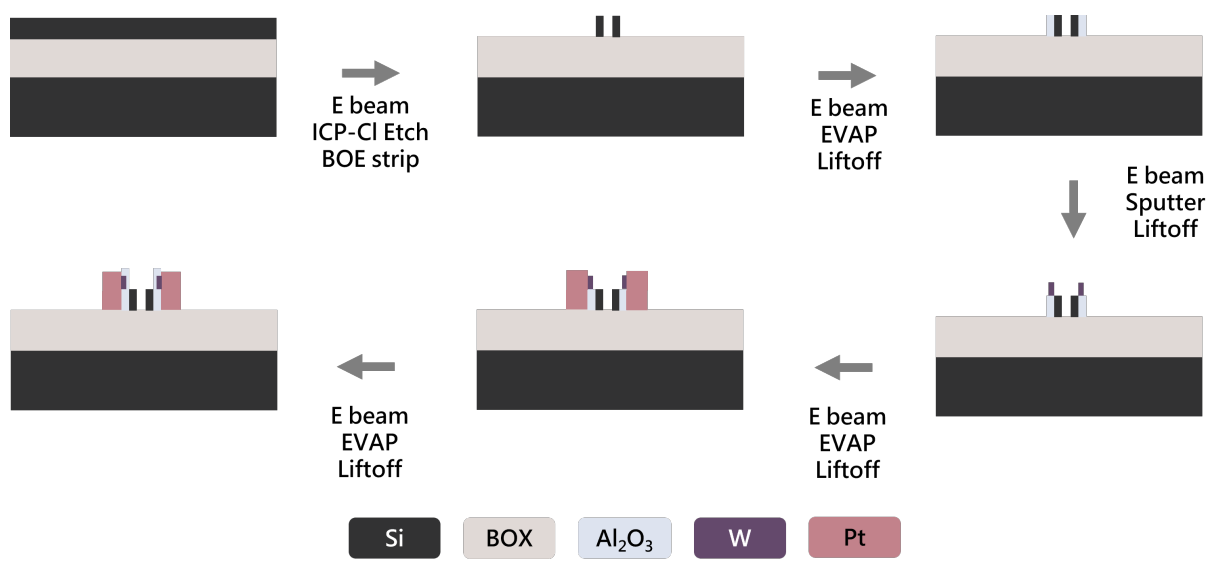

**Fig. S1. Fabrication flow.** For deposition steps positive tone $PMMA$ resist was used for patterning. Bottom $Al_2O_3$ layer is $265nm$ thick and the sputtered tungsten ($W$) layer making up the heating elements is $150nm$ thick. The contact pads are made up of $25nm\ Ti\ /\ 325nm\ Pt$ layers. The final $Al_2O_3$ cladding over the islands is $290nm$ thick.

**Device Characterization:**

The spectrum of the fabricated device was measured via a fiber coupled setup in which the input light was provided by a tunable continuous-wave laser (Santec $TSL-510$) and a low-noise power meter (Keysight $81634B$) was used to collect the output light from the grating couplers. A DAQ (MCC USB 3114) was used to apply the electrical potential profile across the device.

# Supplementary Text

### Section S1: Hamiltonian Tomography:

For a coupled cavity array with $N$ racetrack resonators with grating couplers at first and last sites, the non-interacting system Hamiltonian $H$ can be written as

$$H = \sum_n \mu_n a_n^\dagger a_n + J_n (a_{n+1}^\dagger a_n + a_n^\dagger a_{n+1}) \tag{S2}$$

where $a_n$ denotes onsite destruction operator, $\mu_n$ denotes the onsite potential and $J_n$ denotes the hopping rates between $n^{th}$ and $(n+1)^{th}$ sites. Further using input-output formalism we can write:

$$\begin{aligned}\dot{a}_0 &= -j(\mu_0 a_0 + J_0 a_1) - \frac{\kappa_0}{2} a_0 - \frac{\gamma_0}{2} a_0 - \sqrt{\gamma_0}\, x_0, \\ \dot{a}_n &= -j(\mu_n a_n + J_{n-1} a_{n-1} + J_n a_{n+1}) - \frac{\kappa_n}{2} a_n, \text{where } n = 1,2\ldots N-2 \\ \dot{a}_{N-1} &= -j(\mu_{N-1} a_{N-1} + J_{N-2} a_{N-2}) - \frac{\kappa_{N-1}}{2} a_{N-1} - \frac{\gamma_{N-1}}{2} a_{N-1}\end{aligned} \tag{S2}$$

where $\kappa_n$ denotes the loss rate to environment at each site, $\gamma_n$ denotes the coupling rate to the grating couplers, $x_0$ denotes the destruction operator of the input into the system. The system can be visualized with the aid of the schematic depicted in Fig. S2.

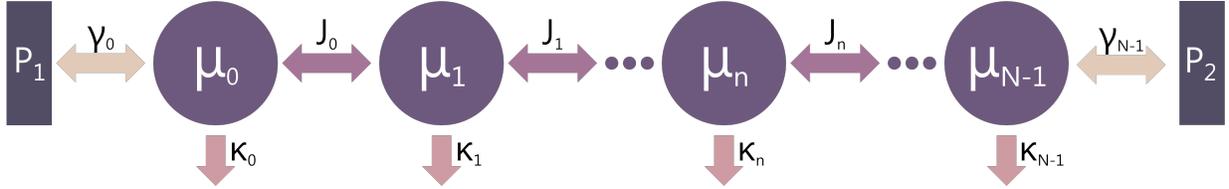

**Fig. S2. Schematic depicting a general CCA device.** Each resonator site is characterized by the on-site potential $\mu_n$ and the loss rate to the environment/absorption is denoted by $\kappa_n$. The hopping rates between the sites are denoted by $J_n$. The array is probed using grating couplers coupled to the first and last sites of the array with coupling rates denoted by $\gamma_n$. The grating couplers are themselves denoted by the ports $P_1, P_2$.

Taking Fourier transform on these equations gives:

$$\begin{aligned}-j\omega a_0 &= -j(\mu_0 a_0 + J_0 a_1) - \frac{\kappa_0}{2} a_0 - \frac{\gamma_0}{2} a_0 - \sqrt{\gamma_0}\, x_0, \\ -j\omega a_n &= -j(\mu_n a_n + J_{n-1} a_{n-1} + J_n a_{n+1}) - \frac{\kappa_n}{2} a_n, \text{where } n = 1,2\ldots N-2 \\ -j\omega a_{N-1} &= -j(\mu_{N-1} a_{N-1} + J_{N-2} a_{N-2}) - \frac{\kappa_{N-1}}{2} a_{N-1} - \frac{\gamma_{N-1}}{2} a_{N-1}\end{aligned} \tag{S3}$$

Rearranging the terms gives:

$$\sqrt{\gamma_0}\, x_0 = j\left(\omega - \left(\mu_0 - j\frac{\kappa_0}{2} - j\frac{\gamma_0}{2}\right)\right) a_0 - j(J_0 a_1),$$

$$0 = j\left(\omega - \left(\mu_n - j\frac{\kappa_n}{2}\right)\right) a_n - j(J_{n-1} a_{n-1} + J_n a_{n+1}), \text{where } n = 1, 2 \ldots N-2 \quad (S4)$$

$$0 = j\left(\omega - \left(\mu_{N-1} - j\frac{\kappa_{N-1}}{2} - j\frac{\gamma_{N-1}}{2}\right)\right) a_{N-1} - j(J_{N-2} a_{N-1})$$

Writing as a matrix:

$$j \begin{pmatrix} \omega - \left(\mu_0 - j\frac{\kappa_0}{2} - j\frac{\gamma_0}{2}\right) & -J_0 & 0 & 0 & \cdots \\ -J_0 & \omega - \left(\mu_n - j\frac{\kappa_n}{2}\right) & -J_1 & 0 & \cdots \\ \vdots & \ddots & \ddots & \ddots & \vdots \\ \cdots & \cdots & 0 & -J_{N-2} & \omega - \left(\mu_{N-1} - j\frac{\kappa_{N-1}}{2} - j\frac{\gamma_{N-1}}{2}\right) \end{pmatrix} \begin{pmatrix} a_0 \\ a_1 \\ \vdots \\ a_{N-1} \end{pmatrix}$$

$$= \begin{pmatrix} \sqrt{\gamma_0} x_0 \\ 0 \\ \vdots \\ 0 \end{pmatrix} \quad (S5)$$

Further we can define $H_{eff}$ as

$$H_{eff} = \begin{pmatrix} \mu_0 - j\frac{\kappa_0}{2} - j\frac{\gamma_0}{2} & J_0 & 0 & 0 & \cdots \\ J_0 & \mu_1 - j\frac{\kappa_1}{2} & J_1 & 0 & \cdots \\ \vdots & \ddots & \ddots & \ddots & \vdots \\ \cdots & \cdots & 0 & J_{N-2} & \mu_{N-1} - j\frac{\kappa_{N-1}}{2} - j\frac{\gamma_{N-1}}{2} \end{pmatrix}, |x_{in}\rangle = \begin{pmatrix} \sqrt{\gamma_0} x_0 \\ 0 \\ \vdots \\ 0 \end{pmatrix} \quad (S6)$$

This allows us to compress Eq. S5 into

$$j(\omega \mathbb{I} - H_{eff})|a_n\rangle = |x_{in}\rangle \quad (S7)$$

Hence, we obtain $|a_n\rangle$ as

$$|a_n\rangle = -j(\omega \mathbb{I} - H_{eff})^{-1}|x_{in}\rangle \quad (S8)$$

From input-output formalism we can also write the output modes as:

$$|y_{out}\rangle = |x_{in}\rangle + |\sqrt{\gamma_n} a_n\rangle \quad (S9)$$

Separately, we know that the two experimentally measurable quantities from the device the reflection spectrum $|R(\omega)|^2$ and the transmission spectrum $|T(\omega)|^2$ of the system can be written as:

$$|R(\omega)|^2 = \left|\frac{y_0}{x_0}\right|^2, |T(\omega)|^2 = \left|\frac{y_{N-1}}{x_0}\right|^2 \tag{S10}$$

Substituting for $y_0$ from Eq. S8, S9 we can obtain $|R(\omega)|^2$ as

$$|R(\omega)|^2 = \left|\frac{y_0}{x_0}\right|^2 = \left|\frac{x_0 - j\gamma_0 x_0 \langle v_0|(\omega\mathbb{I} - H_{eff})^{-1}|v_0\rangle}{x_0}\right|^2 = \left|1 - j\gamma_0 \left\langle v_0\left|\frac{1}{\omega\mathbb{I} - H_{eff}}\right|v_0\right\rangle\right|^2 \tag{S11}$$

where $|v_0\rangle = \begin{pmatrix} 1 \\ 0 \\ \vdots \\ 0 \end{pmatrix}$.

Let $\epsilon_\alpha, |\epsilon_\alpha\rangle$ denote the eigenvalues and eigenvectors of $H_{eff}$, as $H_{eff}$ is complex symmetric matrix we can write

$$H_{eff} = \sum_\alpha \epsilon_\alpha |\epsilon_\alpha\rangle\langle\epsilon_\alpha| \tag{S12}$$

Note that $\langle\epsilon_\alpha|$ denotes the transpose and not the conjugate transpose of $|\epsilon_\alpha\rangle$ and $\langle\epsilon_\beta|\epsilon_\alpha\rangle = \delta_{\beta\alpha}$. This also implies that we can expand $(\omega\mathbb{I} - H_{eff})^{-1}$ as:

$$\frac{1}{\omega\mathbb{I} - H_{eff}} = \sum_\alpha \frac{|\epsilon_\alpha\rangle\langle\epsilon_\alpha|}{\omega - \epsilon_\alpha} \tag{S13}$$

then we can write Eq. S11 as

$$|R(\omega)|^2 = \left|1 - j\gamma_0 \sum_\alpha \langle v_0|\frac{|\epsilon_\alpha\rangle\langle\epsilon_\alpha|}{\omega - \epsilon_\alpha}|v_0\rangle\right|^2 = \left|1 - j\gamma_0 \sum_\alpha \frac{\langle v_0|\epsilon_\alpha\rangle^2}{\omega - \epsilon_\alpha}\right|^2 \tag{S14}$$

Similarly, for transmission spectrum $|T(\omega)|^2$ we have

$$|T(\omega)|^2 = \left|\frac{y_{N-1}}{x_0}\right|^2 = \left|\frac{-j\sqrt{\gamma_0\gamma_{N-1}}\, x_0 \langle v_{N-1}|(\omega\mathbb{I} - H_{eff})^{-1}|v_0\rangle}{x_0}\right|^2$$
$$= \left|-j\sqrt{\gamma_0\gamma_{N-1}} \left\langle v_{N-1}\left|\frac{1}{\omega\mathbb{I} - H_{eff}}\right|v_0\right\rangle\right|^2 \tag{S15}$$

Using Eq. S13 we can write:

$$|T(\omega)|^2 = \left|-j\sqrt{\gamma_0\gamma_{N-1}} \sum_\alpha \langle v_{N-1}|\frac{|\epsilon_\alpha\rangle\langle\epsilon_\alpha|}{\omega - \epsilon_\alpha}|v_0\rangle\right|^2 = \left|-j\sqrt{\gamma_0\gamma_{N-1}} \sum_\alpha \frac{\langle v_{N-1}|\epsilon_\alpha\rangle\langle v_0|\epsilon_\alpha\rangle}{\omega - \epsilon_\alpha}\right|^2 \tag{S16}$$

Going further, exploiting the geometry of the device allows us to write

$$H_{eff}|v_n\rangle = \tilde{\mu}_n|v_n\rangle + J_{n-1}|v_{n-1}\rangle + J_n|v_{n+1}\rangle, \text{ where } \tilde{\mu}_n = \mu_n - j\frac{\kappa_n}{2} - j\frac{\gamma_n}{2};$$

$$|v_n\rangle = \begin{pmatrix} 0 \\ \vdots \\ 1 \\ \vdots \\ 0 \end{pmatrix} n^{th} row \tag{S17}$$

Taking transpose and multiplying from right by $|\epsilon_\alpha\rangle$ we get

$$\langle v_n|H_{eff}|\epsilon_\alpha\rangle = \langle v_n|\tilde{\mu}_n|\epsilon_\alpha\rangle + \langle v_{n-1}|J_{n-1}|\epsilon_\alpha\rangle + \langle v_{n+1}|J_n|\epsilon_\alpha\rangle \tag{S18}$$

$$\Rightarrow \langle v_n|\epsilon_\alpha|\epsilon_\alpha\rangle = \langle v_n|\tilde{\mu}_n|\epsilon_\alpha\rangle + \langle v_{n-1}|J_{n-1}|\epsilon_\alpha\rangle + \langle v_{n+1}|J_n|\epsilon_\alpha\rangle \tag{S19}$$

$$\Rightarrow (\epsilon_\alpha - \tilde{\mu}_n)\langle v_n|\epsilon_\alpha\rangle - J_{n-1}\langle v_{n-1}|\epsilon_\alpha\rangle = J_n\langle v_{n+1}|\epsilon_\alpha\rangle \tag{S20}$$

Also using Eq. S12 we have,

$$\tilde{\mu}_n = \langle v_n|H_{eff}|v_n\rangle = \sum_\alpha \epsilon_\alpha \langle v_n|\epsilon_\alpha\rangle^2 \tag{S21}$$

Now we have all the ingredient equations needed for determining $H_{eff}$. The general method we use for Hamiltonian tomography can be summarized as an algorithm SA1:

---
Tomography algorithm
    (i) Measure reflection spectrum $|R(\omega)|^2$ of the device and fit it to a sum $N$ complex Lorentzian functions to obtain all the $\epsilon_\alpha$'s and $\langle v_0|\epsilon_\alpha\rangle$'s using Eq. S14.
    (ii) For the first site ($n = 0$), obtain the diagonal element $\tilde{\mu}_n$ of $H_{eff}$ using Eq. S21.   (SA1)
    (iii) Normalize Eq. S20 to then obtain off diagonal element $J_n$.
    (iv) Use the value of $J_n$ to obtain all the $\langle v_{n+1}|\epsilon_\alpha\rangle$s needed for the next iteration.
    (v) Repeat steps 3, 4, 5 for $n > 0$ until the entire $H_{eff}$ has been determined.

---

We next demonstrate how the algorithm SA1 looks in practice:
(i) We begin with fitting the experimentally measured reflection spectrum $|R(\omega)|^2$ as sum of $N$ Lorentzians to obtain the values of $\gamma_0, \langle v_0|\epsilon_\alpha\rangle, \epsilon_\alpha$ as

$$|R(\omega)|^2 = \left|1 - j\sum_\alpha \frac{A_\alpha e^{j\phi_\alpha}}{\omega - (\omega_\alpha - j\beta_\alpha)}\right|^2, \text{ where } \epsilon_\alpha = \omega_\alpha - j\beta_\alpha, \quad \gamma_0\langle v_0|\epsilon_\alpha\rangle^2 = A_\alpha e^{j\phi_\alpha} \tag{S22}$$

From normalization it also follows that $\sum_\alpha \langle v_0|\epsilon_\alpha\rangle^2 = 1 \Rightarrow \sum_\alpha A_\alpha e^{j\phi_\alpha} = \gamma_0$. Here, fitting for the complex Lorentzians from the reflection spectrum $|R(\omega)|^2$ is done by minimizing the following expression to ensure that the finally fitted $H_{eff}$ falls in the realm of physical possibility:

$$Min\left\{\| |R(\omega)|^2{}_{predicted} - |R(\omega)|^2{}_{measured} \|^2 + \left|Im\left(\sum_\alpha A_\alpha e^{j\phi_\alpha}\right)\right| + \sum_i |Im(J_i)|\right\} \quad (S23)$$

(ii) For the first site ($n = 0$) we use Eq. S21 to obtain the diagonal element $\tilde{\mu}_0$ as

$$\tilde{\mu}_0 = \sum_\alpha \epsilon_\alpha \langle v_0|\epsilon_\alpha\rangle^2, = \sum_\alpha (\omega_\alpha - j\beta_\alpha) A_\alpha e^{j\phi_\alpha}/\gamma_0 \quad (S24)$$

(iii) Next we determine the first off-diagonal element $J_0$ using Eq. S20 ($(\epsilon_\alpha - \tilde{\mu}_0)\langle v_0|\epsilon_\alpha\rangle = J_0\langle v_1|\epsilon_\alpha\rangle$) for $n = 0$ and normalizing for $\sum_\alpha \langle v_1|\epsilon_\alpha\rangle^2 = 1$, giving:

$$J_0 = \sqrt{\sum_\alpha ((\epsilon_\alpha - \tilde{\mu}_0)\langle v_0|\epsilon_\alpha\rangle)^2} = \sqrt{\sum_\alpha (\omega_\alpha - j\beta_\alpha - \tilde{\mu}_0)^2 A_\alpha e^{j\phi_\alpha}/\gamma_0} \quad (S25)$$

(iv) Finally, we can substitute the obtained $J_0$ back into Eq. S20 to obtain $\langle v_1|\epsilon_\alpha\rangle$s needed for the next iteration as:

$$\langle v_1|\epsilon_\alpha\rangle = \frac{(\epsilon_\alpha - \tilde{\mu}_0)\langle v_0|\epsilon_\alpha\rangle}{J_0} = \frac{(\omega_\alpha - j\beta_\alpha - \tilde{\mu}_0)\sqrt{A_\alpha e^{j\phi_\alpha}/\gamma_0}}{J_0} \quad (S26)$$

(v) Next for $n = 1$, carrying out steps (ii)-(iv) gives:

$$\tilde{\mu}_1 = \sum_\alpha \epsilon_\alpha \langle v_1|\epsilon_\alpha\rangle^2, \quad (\epsilon_\alpha - \tilde{\mu}_1)\langle v_1|\epsilon_\alpha\rangle - J_0\langle v_0|\epsilon_\alpha\rangle = J_1\langle v_2|\epsilon_\alpha\rangle \quad (S27)$$

normalizing for $\sum_\alpha \langle v_2|\epsilon_\alpha\rangle^2 = 1$ gives

$$J_1 = \sqrt{\sum_\alpha ((\epsilon_\alpha - \tilde{\mu}_1)\langle v_1|\epsilon_\alpha\rangle - J_0\langle v_0|\epsilon_\alpha\rangle)^2},$$

$$\text{and } \langle v_2|\epsilon_\alpha\rangle = \frac{(\epsilon_\alpha - \tilde{\mu}_1)\langle v_1|\epsilon_\alpha\rangle - J_0\langle v_0|\epsilon_\alpha\rangle}{J_1} \quad (S28)$$

And so on for $n > 1$ we can keep iterating over the steps (ii)-(iv) until we obtain all the $\tilde{\mu}_n, J_n$ where $n \in [0, \ldots N-1]$. At the end of the algorithm we will have a successfully mapped effective Hamiltonian $H_{eff}$ describing the device.

Couple of things to note in the algorithm are:
1. We assume that all $J_n$ have the same sign, and that it is a valid assumption for our devices.
2. The algorithm SA1 works for determining $H_{eff}$ when we fit the measured reflection spectrum $|R(\omega)|^2$ to a sum of complex Lorentzian functions to obtain the eigenvalues $\epsilon_\alpha$ and spectral weights $\langle v_0|\epsilon_\alpha\rangle$. While we can also fit the transmission spectrum $|T(\omega)|^2$ to a sum of complex Lorentzian functions to obtain the eigenvalues $\epsilon_\alpha$s, as evident form Eq. S16 we cannot uniquely determine the involved spectral weights $\langle v_0|\epsilon_\alpha\rangle, \langle v_{N-1}|\epsilon_\alpha\rangle$ from such a fit.

## Section S2: Thermal crosstalk simulations

For a racetrack resonator we know that $n_{eff} l = m\mu_n^\lambda$, where $n_{eff}$ is the refractive index of the racetrack resonator, $l$ is the length of the resonator, $\mu_n^\lambda$ is the onsite potential (in wavelength units) and $m \in \mathbb{Z}$. Let a segment of length $x$ be affected by change in temperature, such that its refractive index is given by $n(x)$. Resonance condition for the resonator then becomes:

$$n_{eff}(l - x) + \int n(x)dx = m(\mu_n^\lambda + \Delta\mu_n^\lambda) \tag{S29}$$

Removing the constant terms gives

$$\int (n(x) - n_{eff})dx = \int \Delta(n(x))dx = m(\Delta\mu_n^\lambda) \tag{S30}$$

Assuming a constant thermo-optic coefficient $dn/dT$ we can write $\Delta n(x) = \rho \Delta T(x)$. Consequently,

$$\int \rho \Delta T(x) dx = m(\Delta\mu_n^\lambda) \tag{S31}$$

This implies shift in onsite potential $\Delta\mu_n^\lambda \propto \int \Delta T(x) dx$.

Next; to estimate this effect of thermal crosstalk in our system and compare it with typically used thermo-optic (TO) heaters we perform a set of thermal simulations using ANSYS Lumerical HEAT (Fig. S3). In Fig. S3A we have the schematic depicting a conventional TO heater, where the metallic heating element sits directly on top of the resonator segment separated by a uniform and universal $1.5 \mu m\ SiO_2$ cladding. In Fig. S3B we have the island TO heater design we used for our device. For both these cases we have the CCA made up of only 3 sites with the heater placed on the middle site ($n^{th}$). We then record the temperature profile in the shorter straight segment (highlighted in yellow) of the racetrack resonators for the $n^{th}$ and $(n+1)^{th}$ sites as we vary the voltage applied $V_n$ across the heater from $0V$ to $0.8V$. In Fig. S3C we plot the average temperature across the segments given by

$$Avg(T) = \frac{\int T(x)dx}{\int dx} \tag{S32}$$

for neighboring $(n+1)^{th}$ site. Similarly, in Fig. S3D we plot the average temperature across the segment for $n^{th}$ site. It is clear from these plots that even though both heater designs cause a similar increase in the onsite temperature ($n^{th}$), the average temperature in the neighboring sites diverges rapidly between the designs with the temperature difference already greater than 30K at $0.8V$. To study the difference in effects of this thermal crosstalk we then define a dimensionless parameter $\eta$ given by

$$\eta = \frac{\Delta\mu_{n+1}^\lambda}{\Delta\mu_n^\lambda} \tag{S33}$$

Using Eq. S31 we can write the above as

$$\eta = \frac{\Delta\mu_{n+1}^{\lambda}}{\Delta\mu_n^{\lambda}} = \frac{\int \Delta T_{n+1}(x)dx}{\int \Delta T_n(x)dx} = \frac{\int (T_{n+1}(x) - 300)dx}{\int (T_n(x) - 300)dx} \tag{S34}$$

In Fig. S3E we plot the $\eta$ (as %) for both the designs. We can see from the plot that $\eta_{cladded} \sim 0.056$ and $\eta_{island} \sim 0.024$. This implies that the islands TO heaters outperform the typical TO heaters in reducing the effects of thermal crosstalk in the device by more than 50%. In general, we do not expect the $\eta$ values to vary with the voltage $V_n$ (see Section S3) as also verified by the plots.

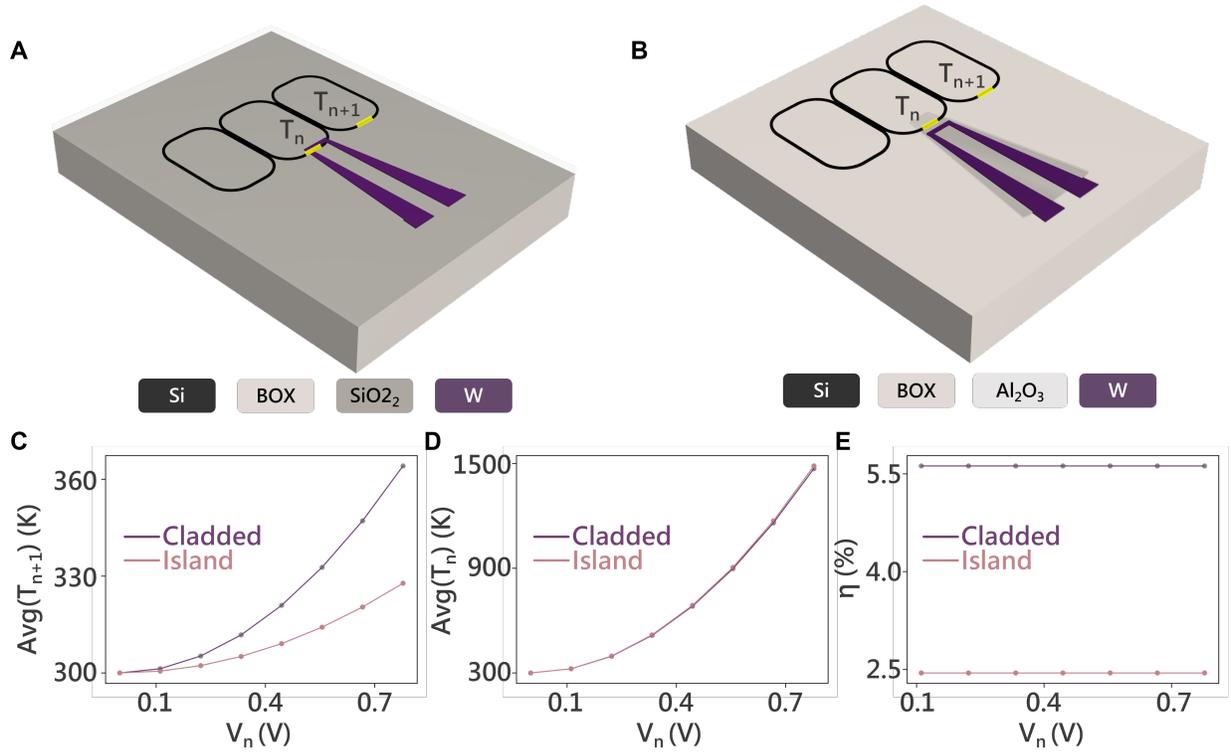

**Fig. S3. Thermal simulation results.** (A) Schematic depicting the device with conventional TO heater sitting directly over the resonator separated by 1.5 $\mu m$ thick oxide cladding (label: Cladded). (B) Schematic depicting a device equipped with the island TO heater used in our paper (label: Island). (C) Plot comparing average temperature across the straight segment (in yellow) of the $(n+1)^{th}$ resonator. (D) Plot comparing average temperature across the straight segment (in yellow) of the $n^{th}$ resonator. (E) Plot comparing $\eta$ factor in % for the two device designs. (purple: cladded, pink: island)

**Section S3: Extracting the thermal crosstalk**

Let voltage $V_n$ be applied to heater $h_n$ at the $n^{th}$ site of the CCA. When only one heater $h_n$ is turned on at time, using Ohmic heating law we can approximate the change in temperature at $n^{th}$ site of the CCA as $Avg(\Delta T_n) \propto V_n^2/R_i$ where $R_n$ is the resistance of heater $h_n$.
For a constant thermo-optic coefficient $dn/dT$, from Eq. S31 we know that $\Delta \mu_n^\lambda \propto Avg(\Delta T_n)$. Combining the above two relations using a proportionality constant $k_1$ allows us to write:

$$\Delta \mu_n^\lambda = k_1 V_n^2 / R_n = \alpha_n V_n^2 \tag{S35}$$

where $\alpha_n = k_1/R_n$.
Since only $h_n$ is turned on, it also follows that for another site $m$ in the array, $\Delta \mu_m^\lambda \propto Avg(\Delta T_n)$. Introducing another set of proportionality constants $\beta_{nm}$ allows us to write:

$$\Delta \mu_m^\lambda = \beta_{nm} \alpha_n V_n^2 = \beta'_{nm} V_n^2 \tag{S36}$$

Using Eq. S33, S34 we can then estimate the thermal crosstalk by fitting for these coefficients $\alpha_n$, $\beta'_{nm}$ by calculating the change in onsite potentials and estimating the eigen-energies of the modified $H_{eff}$ whose initial state $H_{eff}^0$ was fitted with the procedure outlined in Section S1 and demonstrated in the paper.

$$\left[\epsilon_n^\lambda\right]_{estimated} = Eig\left(H_{eff}^0 + [\Delta \mu_n]\mathbb{I}\right), \quad [\Delta \mu_n^\lambda] = [\alpha_n, \beta'_{nm}][V_n^2] \tag{S37}$$

We do this fitting for all the eight heaters $[h_n]$ to obtain corresponding $\alpha_n$s and $\beta'_{nm}$s; accuracy results of which are plotted in Fig. S4. The x-axis denotes the voltage $V_n$ applied to heater $h_n$, the y-axis plots the eigen-energies as wavelengths. The black lines in the background denote the measured locations of eigen-energies on application of $V_n$, the size of the circle denotes the error of the fit from the measured location of eigen-energy. The color of the circles denotes the overall fit error for that particular measurement. The extracted values $\alpha_n$s and $\beta'_{nm}$s are then used to plot Fig. 2 in the paper where we use these to plot the change in $\Delta \mu_n^\lambda$s for all $V_n$s used in the measurements shown below.

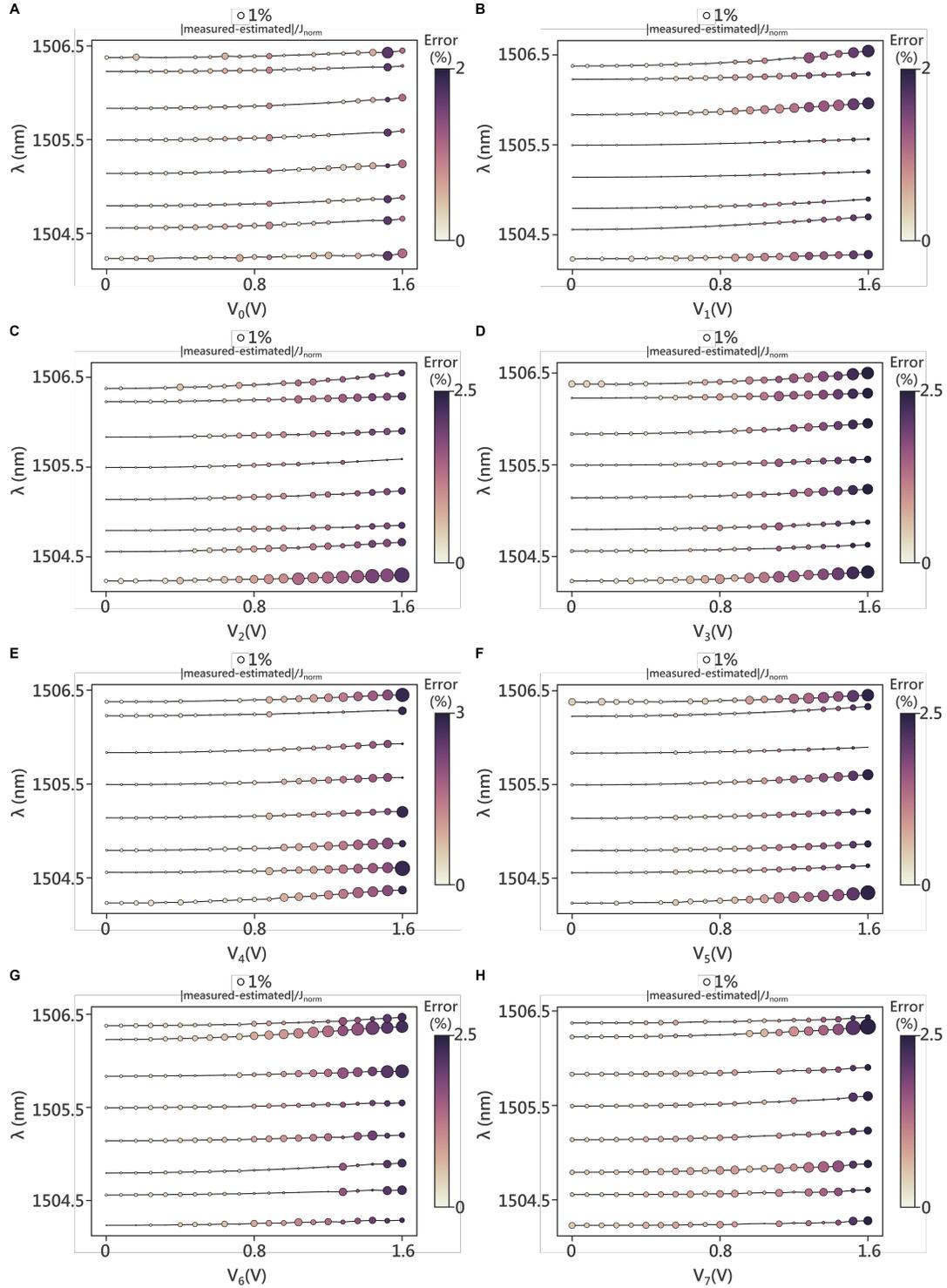

**Fig. S4. Thermal crosstalk extraction.** Fitting accuracy for sweeps with only one heater turned on at a time where the x-axis denotes the voltage applied $V_n$, the y-axis denotes the wavelength with the location of the measured eigen-energies is denoted by the dark black lines in background. The radii of the circles denote the deviation of the fitted value from measured values (scale bar on top). The color of the dots denotes the overall prediction error for that generation.